\documentclass[conference]{IEEEtran}
\IEEEoverridecommandlockouts
\usepackage{cite}
\usepackage{amsmath,amssymb,amsfonts}
\usepackage{algorithm}
\usepackage{algorithmicx}
\usepackage{algpseudocode}
\usepackage{graphicx}
\usepackage{geometry}
\usepackage{textcomp}
\usepackage{color}
\usepackage{xcolor}
\usepackage{hyperref}
\usepackage{listings}
\usepackage{tikz}
\usepackage{multirow}
\usepackage{multicol}
\usepackage{makecell}
\usepackage{amsmath}
\usepackage{booktabs}
\usepackage{utfsym}
\usepackage{wrapfig}
\usetikzlibrary{shapes,arrows}
\usetikzlibrary{shapes.geometric, arrows}
\def\BibTeX{{\rm B\kern-.05em{\sc i\kern-.025em b}\kern-.08em
    T\kern-.1667em\lower.7ex\hbox{E}\kern-.125emX}}

\makeatletter
\newcommand{\linebreakand}{%
  \end{@IEEEauthorhalign}
  \hfill\mbox{}\par
  \mbox{}\hfill\begin{@IEEEauthorhalign}
}
\makeatother

\begin{document}
\title{\Large\bf$TIUP$: Effective Processor Verification with Tautology-Induced Universal Properties\\
\thanks{*Corresponding author}
}


\author{\IEEEauthorblockN{1\textsuperscript{st} Yufeng Li}
\IEEEauthorblockA{\textit{Institute of Software, Chinese Academy of Sciences} \\
\textit{University of Chinese Academy of Sciences}\\
Beijing, China \\
e-mail: yufeng@nfs.iscas.ac.cn}
\and
\IEEEauthorblockN{2\textsuperscript{nd} Yiwei Ci}
\IEEEauthorblockA{\textit{Institute of Software, Chinese Academy of Sciences} \\
\textit{University of Chinese Academy of Sciences}\\
Beijing, China \\
e-mail: yiwei@iscas.ac.cn}
\and
\IEEEauthorblockN{3\textsuperscript{rd} Qiusong Yang$^{\ast}$}
\IEEEauthorblockA{\textit{Institute of Software, Chinese Academy of Sciences} \\
\textit{University of Chinese Academy of Sciences}\\
Beijing, China \\
e-mail: qiusong@iscas.ac.cn}
}

\maketitle

\begin{abstract}
Design verification is a complex and costly task, especially for large and intricate processor projects. Formal verification techniques provide advantages by thoroughly examining design behaviors, but they require extensive labor and expertise in property formulation. Recent research focuses on verifying designs using the self-consistency universal property, reducing verification difficulty as it is design-independent. However, the single self-consistency property faces false positives and scalability issues due to exponential state space growth. To tackle these challenges, this paper introduces $TIUP$, a technique using tautologies as universal properties. We show how $TIUP$ effectively uses tautologies as abstract specifications, covering processor data and control paths. $TIUP$ simplifies and streamlines verification for engineers, enabling efficient formal processor verification.
\end{abstract}


\section{Introduction}
Verification is crucial in processor design and manufacturing\cite{lonsing2019unlocking}, with modern processors having optimization components like branch predictors, prefetchers, and out-of-order execution units that make verification expensive. Simulation-based verification involves many test cases and an oracle to compare instruction execution results. In contrast, formal verification, like model checking\cite{clarke1999jr,clarke2018model}, constructs a mathematical representation of the system and proves it adheres to prescribed properties. It excels at discovering corner cases\cite{reid2016end}, but specifying the properties of the design under verification (DUV) is challenging, the quality of verification heavily depends on expertise in property formulation\cite{bormann2007complete, DBLP:conf/charme/KatzGG99}.

To simplify property formulation, ISA-Formal\cite{reid2016end} proposed an end-to-end approach for formal processor verification. It automates architectural specification language (ASL) into properties for each instruction. This sparked research on ISA-level verification for RISC-V processors (RISCV-Formal\cite{riscv-formal}) and ILA\cite{subramanyan2015template,huang2018instruction,huang2019ilang} for accelerators. Defining properties for individual instructions is a non-trivial task\cite{reid2016trustworthy}. Moreover, instructions are usually verified independently, and the modeling of inter-instruction interactions is insufficient. $QED$ (Quick Error Detection) proposes a design-independent universal property called self-consistency, based on the idea that a processor executes two identical operations, the result should be consistent\cite{DavidLin2014EffectivePV}. $SQED$\cite{DavidLin2015ASA,singh2018logic,lonsing2019unlocking} and $S^2QED$\cite{MohammadRahmaniFadiheh2018SymbolicQE} are its symbolic execution versions. However, relying solely on the self-consistency property has issues. Instruction enumeration becomes complex and time-consuming, especially for bugs with long activation sequences\cite{ganesan2021effective}. Additionally, self-consistency does not address single-instruction bugs. $C\text{-}S^2QED$\cite{devarajegowda2020gap} extends $S^2QED$ to target single-instruction bugs by generating a comprehensive set of properties using metamodeling techniques\cite{devarajegowda2018meta}. Unfortunately, $C\text{-}S^2QED$ properties are no longer universal because of the requirement to formally specify the semantics of individual instruction.

In this paper, we propose an end-to-end formal verification approach for processors based on tautology-induced universal properties ($TIUP$). Instead of relying on a single universal property, $TIUP$ explores a set of universal properties present in functional units. The fundamental properties, referred to as seeds, focus on data paths and simple control logic. Proposition tautology templates are replaced with seed formulas to generate tautologies covering complex control paths. By symbolically executing instruction sequences related to properties through model-checking engines, we can then verify whether the results of the instruction execution conform to tautology specifications. Unlike $SQED$ and variants, $TIUP$ verifies multiple properties to eliminate missed detection. For instance, consider the associative law $x-y-z=x-(y+z)$ in first-order logic (FOL) with arithmetic. It should hold in any processor that implements the addition and subtract operations, serving as one of the essential seed properties. These seed properties play a pivotal role in focusing on the data paths and simple control logic relevant to processor verification. By replacing elements $P$ or $Q$ in the proposition tautology templates $(P \&\& Q) \to P$ with the aforementioned associative law, we can generate a new tautology that covers jump and logic computation. After correctly executing instruction sequences from tautology, the value in the result register should be 1 (true). In $x-y-z=x-(y+z)$, two distinct sub-instruction sequences are generated from $x-y-z$ and $x-(y+z)$, respectively. The target result register serves as an indicator of whether they yield identical outcomes. In comparison to $SQED$ and variants, our approach overcomes the inherent vulnerability in repetitive comparisons, which are susceptible to identical flaws.    

In summary, the contributions of this paper are the following:
\begin{itemize}
    \item We propose the use of universal properties derived from FOL tautologies that are independent of microarchitectural details for processor verification; 
    \item We design and implement $TIUP$ that can formally verify the implementation of a processor based on a set of universal properties;
    \item We evaluate the effectiveness and performance of $TIUP$ on two real open-source processors (in-order and out-of-order) based on the RISC-V architecture. $TIUP$ can effectively detect anomalies related to both the data path and control path of processors.
\end{itemize}

The rest of this paper is structured as follows. Section \ref{sec: motivation} provides additional details regarding the motivation behind this study, demonstrating the necessity of extending the universal property. Section \ref{sec: design} elaborates on $TIUP$ design. We first introduce an overview and formal model of $TIUP$ and then proceed to illustrate each component of the design. Section \ref{sec: evaluation} presents the results of our experimental evaluation of the proposed approach. Finally, in Section \ref{sec: conclusion}, we conclude this paper.

\section{Motivation}\label{sec: motivation} 
The universal property is inherently design-independent, meaning that no manual property formulation is required for verification purposes\cite{lonsing2019unlocking}. $QED$ and variants leverage the idea of instruction execution self-consistency to formulate a single universal property. The self-consistency property specifies that when both the original and duplicate instruction sequences are executed from a $QED\text{-}consistent$ state (i.e., registers and memory locations that are partitioned separately for the original and duplicate instruction sequences hold identical values), the resulting state must also be $QED\text{-}consistent$. 

\subsection{False Positives with Self-consistency}
As mentioned earlier, both the original and duplicate sequences can be affected by identical flaws, resulting in false positives during self-consistency checks. For instance, consider Listing \ref{lst: false positive}, where both the original and duplicate instruction sequences contain multiple \textit{ADD} instructions that are mistakenly interpreted as \textit{SUB} instructions. Despite this discrepancy, the self-consistency property remains intact, making it impossible for the solver to identify a counterexample. 

A more intricate case arises in the detection of control flow bugs. If both the original and duplicate instruction sequences have incorrect branch directions, the self-consistency check alone cannot identify the bug. The enhanced EDDI-V approach\cite{singh2019symbolic} is designed to detect incorrect branch directions. As demonstrated in Listing \ref{lst: enhanced EDDI-V}, it captures the target address of each control flow instruction (branch or jump), such as ``2'' and ``5''. If the original sequence jumps correctly (``2''), and the duplicate sequence jumps erroneously (``5''), a non-$QED$ instruction leads to self-consistency being violated. However, there is also a scenario where the original and duplicate sequences mistakenly jump to ``5'', resulting in false positives.

To eliminate missed detections caused by false positives, researchers have supplemented self-consistency by specifying the semantics of all individual instruction\cite{lonsing2019unlocking, devarajegowda2020gap}. However, these measures compromise the advantages of design-independent universal properties. For $TIUP$, the determination of correctness is based on the requirement that the execution of the processor must meet all tautologies, thus false positives will not appear when checking all universal properties. For instance, during the verification of the property $x-y-z=x-(y+z)$, if ``$+$'' is mistakenly decoded as ``$-$'' in the implementation, the solver can promptly give a counterexample as $x-y-z \ne x-(y-z)$.
\lstset{
	keywordstyle=\color[RGB]{255,99,71},
	keywords={},
	frame=single,
	xleftmargin=0.5em,
	xrightmargin=0.5em,
	captionpos=b,
	breaklines=true,
	basicstyle=\scriptsize,
	caption = {False positive for self-consistency}
}
\begin{lstlisting}[mathescape,float,label={lst: false positive}]
//Original instruction sequence
$R4=R1+R2 \qquad \Rightarrow \qquad R4=R1-R2$ 
$R5=R4+R3 \qquad \Rightarrow \qquad R5=R4-R3$

//Duplicate instruction sequence
$\boldsymbol{R20=R17+R18 \qquad \Rightarrow \qquad R20=R17-R18}$
$\boldsymbol{R21=R20+R19 \qquad \Rightarrow \qquad R21=R20-R19}$

$BNE \quad R4,\boldsymbol{R20} \quad ERROR\_DETECTED$
\end{lstlisting} 
\lstset{
	keywordstyle=\color[RGB]{255,99,71},
	keywords={}, 
	frame=single,
	xleftmargin=0.5em,
	xrightmargin=0.5em,
	captionpos=b,
	breaklines=true,
	basicstyle=\scriptsize,
	caption = {Enhanced EDDI-V: erroneous branch direction}
}
\begin{lstlisting}[mathescape,float,label={lst: enhanced EDDI-V}]
$\boldsymbol{Original \quad sequence  \quad \qquad \qquad \qquad \quad   Duplicate \quad sequence}$
$R0=0, \quad R4=1, \quad R5=2 \qquad  \qquad \qquad  R16=0, \quad R20=1, \quad R21=2$ 
$0: R3=R5-R4 \qquad \qquad \qquad \qquad \qquad \qquad 0: R19=R21-R20$
$1: BEQ \quad R3,R0 \quad \#5$ $\qquad \qquad \qquad \qquad \quad 1: BEQ \quad R19,R16 \quad \#5$
   //$R3 \quad != 0$                                 //$R19 \quad != 0$
   //$Branch \quad not \quad taken   \qquad \qquad  \qquad  // \boldsymbol{Error: Branch \quad taken}$
$2: \boldsymbol{R2=R4+R5}  \qquad  \qquad \qquad \qquad \qquad  //Ignore \quad QED \quad instruction$ 
                                  $(2: R18=R20+R21)$
   //$\boldsymbol{R2=3}$          

                        //$\boldsymbol{Use \quad Non\text{-}QED \quad instruction}$
                                   $\boldsymbol{5: lui \quad R18, \quad 0}$ 
                                     //$\boldsymbol{R18=0}$
                             
\end{lstlisting}

\subsection{Under-constrained State Space with Self-consistency}
Additionally, to verify the self-consistency property, it is necessary to make a copy of each original instruction sequence and enumerate various combinations of instructions. As the instruction sequence increases, the solver will time out in the face of an exponentially expanding state space. For BMC, deep unrolling may be prohibitive\cite{ganesan2021effective}. Although $S^2QED$\cite{MohammadRahmaniFadiheh2018SymbolicQE} reduces the instruction sequence by instantiating two processors, it has limitations in terms of false positives and larger design sizes due to two processors.

$TIUP$ verifies multiple universal properties, which is equivalent to a more detailed decomposition of the self-consistency property, thus reducing the state space that needs to be proven for the system.

\section{Effective Processor Verification with Tautology-Induced Universal Properties (\emph{TIUP})}\label{sec: design}
\begin{figure}[tp]
\centerline{\includegraphics[width=0.50\textwidth]{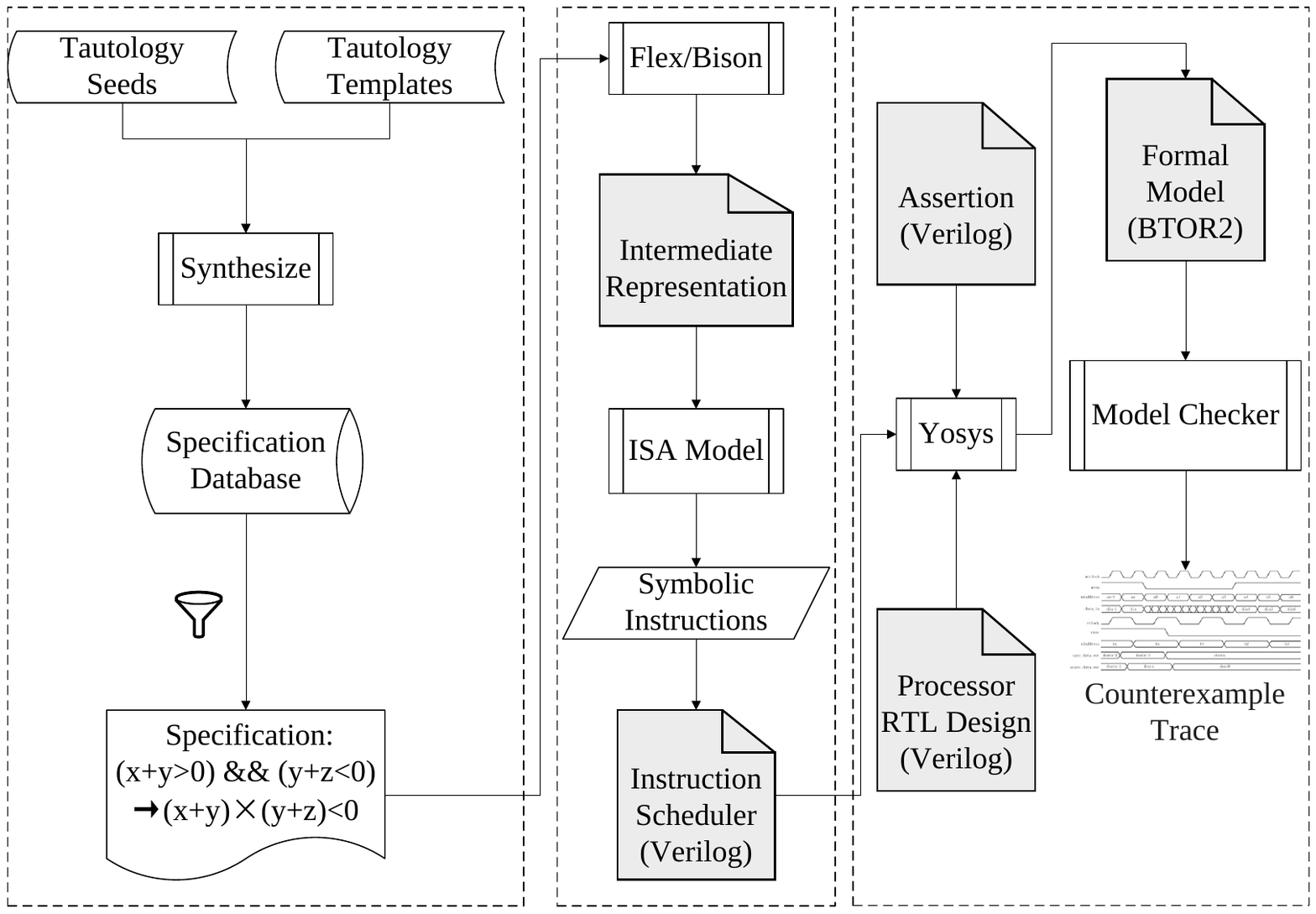}}
\caption{Workflow of $TIUP$}
\label{fig: workflow}
\end{figure}

$TIUP$ is an end-to-end formal verification approach that employs tautologies as abstract specifications. Fig. \ref{fig: workflow} illustrates the workflow of $TIUP$.

\subsection{Overview}\label{sec: overview}
Specifications define the anticipated behaviors of the processor. In end-to-end verification, instructions serve as the interface between the specifications and the implementations of the processor.\\ 
\textbf{Abstract specification.} To verify the correctness of target hardware, one option is to specify its properties and identify at which level of abstraction these properties can be evaluated for truth or falsity \cite{clarke2018model}. Tautology is a formalism that can specify the validity of the processor. An abstract specification is expressed as follows:
\begin{align}
    \vDash f
\end{align}
$\vDash f$\cite{weisstein} refers to $f$ as a tautology which is always true.\\  
\textbf{Formal model.}\label{sec: formal model} 
A processor can be conceptualized as a state transition system: $\langle S, S_{0}, r, i \rangle$, where $S$ represents a vector of state variables, and $S'$ denotes primed variables in the next state; $S_{0}$ is a predicate that denotes initial states; $r$ denotes the transition relation, and $r(S, S', i)$ indicates that the system transitions from $S$ to $S'$ after the instruction $i$ is executed. For a safety property $SP$, the verification process is to check if the transition system will ever reach a violating state $S_{v}$ in which $SP$ does not hold. $Inv$ refers to an invariant. The verification model is presented as follows:
\begin{align}
    S_{0}(S) &\to Inv(S) \label{eq: S0} \\
    Inv(S) \&\& r(S, S', i) &\to Inv(S') \label{eq: transition}\\
    Inv(S) &\to SP \label{eq: proof}
\end{align}
(\ref{eq: S0}) indicates that $Inv(S)$ holds in the initial state; (\ref{eq: transition}) indicates that the invariant should hold in the next state if it holds in the previous one; (\ref{eq: proof}) indicates that $Inv(S)$ implies $SP$, reflecting the fact that $SP$ holds in the whole system. For an abstract specification $\vDash f$, it can map to an instruction sequence $I_{f}$ (refer to \ref{sec: IR}). The meaning behind a tautology violation in a transition system can be formally defined as follows:
\begin{align}
    S_{0}(S) \&\& I_{f}=C(f) \&\& r(S_{0}(S), S_{v}, I_{f}) \&\& !SP \to \bot \label{eq: counterexample}
\end{align}
$I_{f}=C(f)$ represents that mapping of tautology $f$ to instruction sequence $I_{f}$. (\ref{eq: counterexample}) asserts that if the transition system moves from a safety state $S_{0}(S)$ to a violating state $S_{v}$ while following the instruction sequence specified by the tautology $f$, the system fails to meet the requirements of the abstract specification $\vDash f$.\\  
\textbf{Establishing abstract specifications.} The objective of $TIUP$ is to perform formal verification to determine if a processor violates a given set of abstract specifications. Therefore, as a prerequisite for the verification process, it is essential to establish a well-defined set of abstract specifications. In $TIUP$, a synthesis methodology is employed to establish this set of abstract specifications.\\
\textbf{Mapping abstract specifications.} The specifications determine the instructions under verification (IUVs). $TIUP$ generates a symbolic instruction sequence using a specific tautology, with the aim of finding a mapping that leads the transition system from the initial state $S_{0}(S)$ to the violating state $S_{v}$ through the instruction sequence $I_{f}=\{i_{0}, i_{1}, ..., i_{m}\}$. This process involves lexical and syntax analysis of the tautology, transforming it into an ISA-independent Intermediate Representation (IR). The IR is then further transformed into a symbolic instruction sequence, which is stored in a scheduler module. The scheduler module is integrated into the processor's fetch stage for verification purposes and is not included in the final manufactured IC.\\
\textbf{Interfacing RTL design and model checking.} To perform formal verification of the processor's RTL using formal tools, the properties to be verified need to be expressed as assertions and integrated into the RTL code. Subsequently, the entire RTL is transformed into a file format compatible with model-checking tools.

\subsection{Establishing Abstract Specifications}\label{sec: synthesizing}
An abstract specification is created by several seeds and a template. Seeds are first-order tautologies that encompass universal properties pertaining to processors' basic functions. Instructions in processors can be classified into several categories, such as arithmetic and logic, memory access, and control. For arithmetic and logic, the seeds include associative law $\big(x-y-z=x-(y+z)\big)$, De Morgan's theorem for Boolean algebra $\Big(x \oplus y = \sim \big((x \& y) | (\sim x \& \sim y)\big)\Big)$, etc. For memory access, the tautology $ld\big(st(mem, j,v), j\big)=v$ is used as an abstract specification. The function $st$ takes memory $mem$, index $j$, and value $v$ as inputs, returning a memory variable. The function $ld$ receives a memory and an index as its parameters and returns the value fetched from memory. For control, $TIUP$ utilizes tautologies that contain implication connectives that can express conditional jumps.

To cover the complex control logic of processors, abstract specifications need to describe complex control flow. $TIUP$ utilizes template synthesis to construct such specifications. Templates are propositional tautologies, where the value of propositional is true or false, for example, implication introduction $\big((P \to Q) \to (!P || Q)\big)$. Non-logical connective elements (e.g. $P$,$Q$) of a template can be replaced by seeds through a process, called instantiation. During the replacement process, multiple occurrences of a non-logical connective are replaced with the same seed, thus each instantiated template is still a tautology.

Algorithm \ref{alg: SynthesizeTautology} adopts a depth-first search strategy to instantiate templates. It takes a set of tautology templates, denoted as $F$, and a set of tautology seeds, denoted as $E$, as its inputs. A template is selected from the set and instantiated in sequence (line 2). $V$ keeps non-logical connectives of a template formula (line 3). Seeds are selected sequentially from the seed set ($E$) to replace elements in $V$. Once the replacement is completed, an instance of $f$ is obtained, followed by backtracking to the previous element in $V$ (replace $P$ before replacing $Q$). Finally, the output of Algorithm \ref{alg: SynthesizeTautology} is a collection of instantiated tautologies.
\begin{algorithm}[tp]
    \footnotesize
    \caption{SynthesizeTautology}
    \label{alg: SynthesizeTautology}
    \renewcommand{\algorithmicrequire}{\textbf{Input:}}
    \renewcommand{\algorithmicensure}{\textbf{Output:}}

    \begin{algorithmic}[1]
        \Require $F$ is a tautology template set, $E$ is a tautology seed set
        \Ensure $G$ is a collection of instantiated tautology templates
        \State $G \gets \emptyset$;
        \For{$f \in F$}
            \State $V \gets \emptyset$;\Comment{$V$ keeps non-logical connectives}
            \State $N \gets 0$;
            \For{$o \in f$}
                \If{$o$ is non-logical connective $\&\&$ $o \notin V$}
                    \State $V \gets V \cup \{o\}$;
                    \State $N \gets N+1$;
                \EndIf
            \EndFor
            \State $stack \gets \emptyset$;
            \State $stack.push\big((0,f)\big)$;
            \While{$!stack.empty()$}
                \State $(m,rf)=stack.top()$; $stack.pop()$;
                \If{$m==N$}\Comment{all non-logical connectives have been replaced}
                    \State $G \gets G \cup \{rf\}$;\Comment{$rf$ is an instance of $f$}
                    \State continue;
                \EndIf
                \For{$e \in E$}
                    \State replace element $V[m]$ in $rf$ with $e$;
                    \State $stack.push\big((m+1,rf)\big)$;
                \EndFor
            \EndWhile
        \EndFor
    \end{algorithmic}
\end{algorithm}

Depending on the complexity of the processor or computational resource constraints, these universal properties can be verified separately, which can alleviate the state explosion problem encountered during search in comparison with a single self-consistency property.

\subsection{Mapping Abstract Specifications}
The instantiated tautologies are leveraged to map into instructions under verification $\big(I_{f}=C(f)\big)$, which can be utilized to determine whether the tautologies hold. Firstly, an IR can be produced via lexical and syntax analysis of the instantiated tautologies. Secondly, the IR is transformed into a specific instruction sequence according to the ISA model. These symbolic instructions are then stored in a scheduler connected to the pipeline's fetch stage.

\subsubsection{Intermediate representation}\label{sec: IR}
$TIUP$ compiles an instantiated tautology into an IR that is ISA-independent according to the results of lexical and syntax analysis. IR comprises a sequence of directives that describe computations and data flow transformations within the tautology. In this representation, constant symbols and variables are represented as operands, whereas predicates, functions, and connectives are translated into opcodes. The IR for $\vDash (x+y>0)\&\&(y+z<0)\to(x+y)\times(y+z)<0$ is shown in Listing \ref{lst: IR}. The logical implication describes a control-related specification. The specification $\vDash (x+y>0)\&\&(y+z<0)\to(x+y)\times(y+z)<0$ describes that when the execution result of antecedent $\big((x+y>0)\&\&(y+z<0)\big)$ is true, the control flow jumps to the instruction block of consequent $\big((x+y)\times(y+z)<0\big)$ and the execution result of consequent must be true. The result of each tautology is accumulated through the \textit{and} operation and written back the updated result to a special register $\big(\%result\_reg = and$ $i32$ $\%tmp7,$ $\%result\_reg\big)$. A well-defined IR enables the ease of employing $TIUP$ for the verification of processors with different ISAs. IR is transformed into specific instruction sequences according to the ISA model, for example, under the RV32I model, the directive $\%tmp1 = add$ $i32$ $\%x,$ $\%y$ is transformed into a 32-bit addition instruction, where the opcode is $0110011$ ($add$), two source registers are $\%x$ and $\%y$, destination register is $\%tmp1$.
\lstset{
	keywordstyle=\color[RGB]{255,99,71},
	keywords={},
	frame=single,
	xleftmargin=0.5em,
	xrightmargin=0.5em,
	captionpos=b,
	breaklines=true,
	basicstyle=\scriptsize,
	caption = {IR of $\vDash (x+y>0)\&\&(y+z<0)\to(x+y)\times(y+z)<0$}
}
\begin{lstlisting}[mathescape,float,label={lst: IR}]
%tmp1 = add i32 %x, %y                    |        +       
%tmp2 = sgt i32 %tmp1, 0                  |        > 
%tmp3 = add i32 %y, %z                    |        +
%tmp4 = slt i32 %tmp3, 0                  |        <
%tmp5 = logic_and i32 %tmp2, %tmp4        |        $\&\&$
beq %tmp5 1, label %if, label %else       |        $\to$
                                          | 
if:                                       |  
%tmp6 = mul i32 %tmp1, %tmp3              |        $\times$
%tmp7 = slt i32 %tmp6, 0                  |        <
%result_reg = and i32 %tmp7, %result_reg  |        $\&$
                                          |
else:                                     |
other instruction blocks                  |
\end{lstlisting}

\subsubsection{Scheduler}\label{sec: scheduler}
The IUVs mapped from tautologies are stored in a scheduler module, which is attached to the instruction fetch stage of the pipeline. These IUVs are dispatched to the pipeline according to the fetch instruction logic. The scheduler stores three types of instructions: IUVs, $Finish\_Reg \gets 1$, $Result\_Reg \gets 0$. $Finish\_Reg$ and $Result\_Reg$ are two special registers. $Result\_Reg$ is initially set to $1$ and used to preserve the final result. After all the instruction sequences corresponding to a tautology are committed, the satisfaction of that tautology is written back to $Result\_Reg$. Once the processor violates a tautology, the value of $Result\_Reg$ will be set to $0$. The instruction $Finish\_Reg \gets 1$ flags that all IUVs have been committed. 

\begin{figure}[tp]
\centerline{\includegraphics[width=0.45\textwidth]{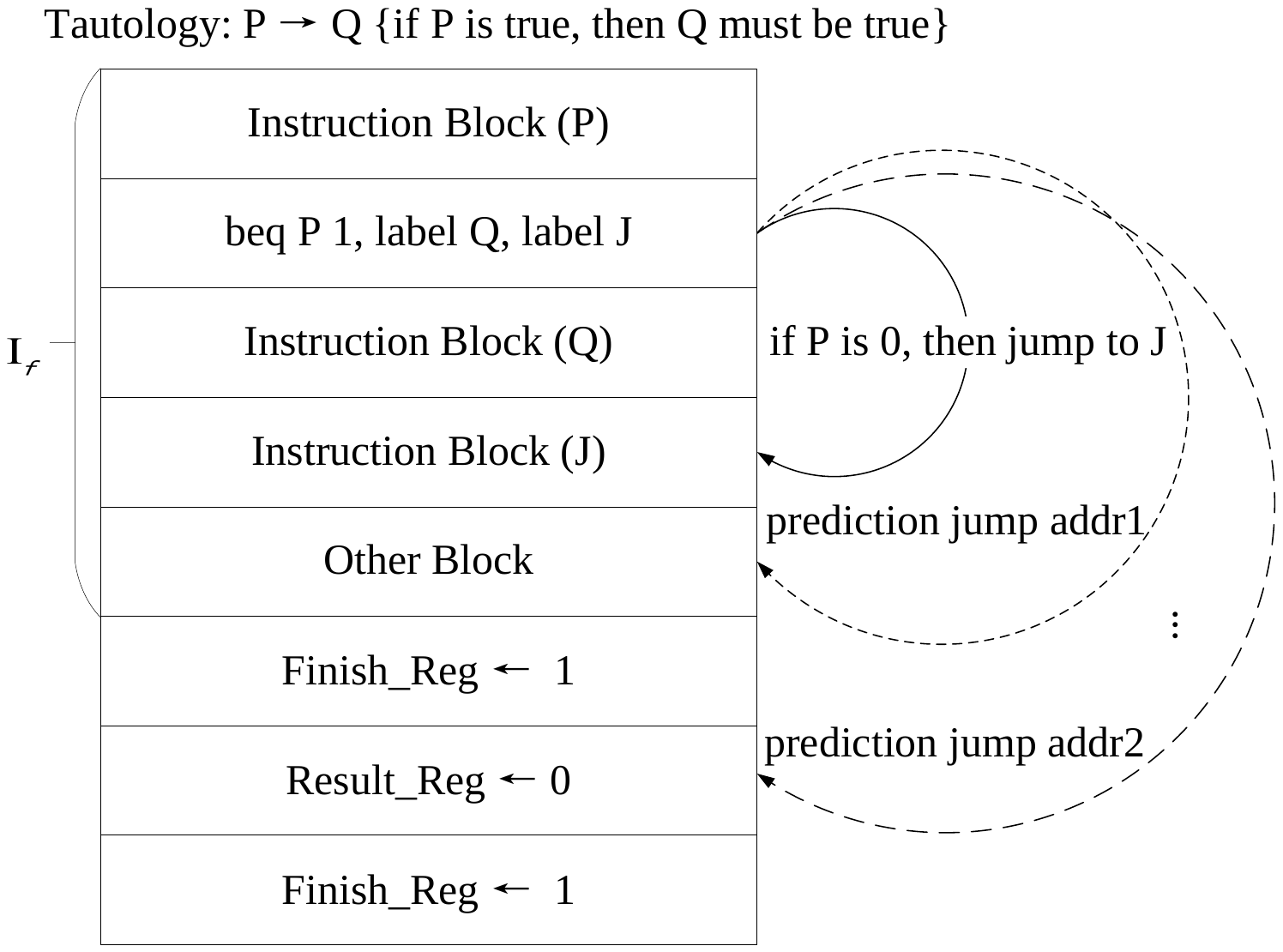}}
\caption{Dispatch instructions from scheduler}
\label{fig: controlflow}
\end{figure}

$TIUP$ is able to verify control flow. As depicted in Fig. \ref{fig: controlflow}, a part of IUVs stored in the scheduler derived from tautology $\vDash P \to Q$. If $P$ is false, the branch is expected to jump to $J$ (the exact offset is calculated in Section \ref{sec: IR}), but when the branch jumps to other instruction blocks or no jump occurs, the value of register $Result\_Reg$ will be set to $0$. If the range of the jump exceeds the length of the instruction queue in the scheduler, the instruction $Result\_Reg \gets 0$ is dispatched to indicate an error in the branch prediction unit. 

\subsection{Interfacing RTL Design and Model Checking}\label{sec: interfacing}
In this phase, the scheduler (see Section \ref{sec: scheduler}) and assertion modules are integrated into the DUV's RTL. Assertions are functions that evaluate a boolean value over the state-holding elements of the processor. The safety property ($SP$, refer to Section \ref{sec: formal model}) expressed as an assertion is inserted into the RTL. A feature of $TIUP$ is its ability to reduce the burden of engineers in writing a plethora of tedious and error-prone assertions. The safety property to be verified is as follows:
\begin{equation}\label{eq: property}
    Finish\_Reg \to Result\_Reg
\end{equation}
Upon completion of the execution and writeback of the instruction sequence under the mapping of tautologies, the register recording the results should be set to 1.

$TIUP$ converts RTL code into a word-level model checking format - BTOR2\cite{niemetz2018btor2} by the open synthesis suite Yosys\cite{wolf2016yosys}. The model-checking tool is able to find a counterexample quickly if the DUV fails to comply with the required tautology.

\section{Evaluation}\label{sec: evaluation}
We evaluate $TIUP$ to assess its efficacy and practicality across two CPU designs. BIRISCV\cite{biriscv} is a dual-issue, seven-stage pipeline in-order processor, whilst RIDECORE\cite{ridecore} is a dual-issue, six-stage pipeline out-of-order processor. We injected 20 distinct logic anomaly scenarios from\cite{lonsing2019unlocking, singh2018logic, DavidLin2014EffectivePV, zhang2018end} as well as some real bug scenarios into the two CPUs' RTL. Model checking was performed using the open-source tool Pono\cite{DBLP:conf/cav/MannILYZBGB20} on an Intel(R) Core(TM) i9-10900K CPU with 3.70GHz and 64GB of RAM. 

We compared $TIUP$ with $SQED$ and $S^2QED$, both of which are based on a universal property. In comparison with the original $SQED$, we only implemented the basic EDDI-V transformation, which targets self-consistency property checking. BMC was utilized as the proof mode for all three methods.

The results are summarized in Table \ref{tab: anomaly}. \textit{Category} refers to whether the anomaly is related to single-instruction or multiple-instruction. The state of being independent of previously executed instructions pertains to single-instruction anomalies, while multiple-instruction anomalies require specific instruction sequences to trigger that error scenario. \textit{Processor} refers to the distribution of anomalies across the two processors. The anomalies $a01$ and $a02$ are not artificially injected into the processors. For the anomaly $a01$, $TIUP$ identifies that BIRISCV does not support a division-by-zero exception. While division-by-zero exceptions are not mandatory for processor designs, they are generally used to ensure program stability and correctness. On the other hand, $TIUP$ identifies that RIDECORE, as an incomplete demo, does not include division and modulo instructions because it only implements a subset of the RV32I. 

The following observations can be made based on Table \ref{tab: anomaly}:

\textbf{Observation 1:} As discussed in Section \ref{sec: motivation}, self-consistency can result in false positives. Because of the inherent limitations of the self-consistency property, $SQED$ and $S^2QED$ missed single-instruction anomalies, whereas $TIUP$ can detect both. 

\textbf{Observation 2:} $SQED$ and $S^2QED$ are capable of efficiently detecting some bugs under the length of two instructions. The shortest computation spent of $TIUP$ to detect a bug is longer than $SQED$ and $S^2QED$ because the property (\ref{eq: property}) is only checked after all instructions corresponding to specified tautology have been committed. However, the longest computation time to detect a bug is shorter in $TIUP$ than in $SQED$ and $S^2QED$. $TIUP$ can verify all tautology specifications simultaneously, meaning that multiple tautology specifications may cover a bug, and we select the shortest bug-finding time. This is because the solver's search object is constrained by each tautology, thereby resulting in a shorter computation time than $SQED$ and $S^2QED$ (Table \ref{tab: anomaly} shows the longest computation for detecting an anomaly in $TIUP$ was less than 1.5h of $SQED$ and $S^2QED$). Since the state space substantially influences the complexity of the proof, being able to divide the search space makes $TIUP$ more scalable for large-scale designs. 

\textbf{Observation 3:} $TIUP$ was unable to detect the anomaly $a17$, where the source operand of an instruction is misidentified as 0. This anomaly, however, does not violate any of the tautologies verified in our experiments. In fact, our future research will focus on identifying the minimal set of universal properties that can comprehensively cover the behavior of the processor. 
\begin{table*}[tbhp]
    \caption{Anomaly detection results}
    \begin{center}
        \resizebox{\linewidth}{!}{
        \begin{tabular}{|c|l|c|c|c|c|c|c|}
            \hline 
            \multirow{2}{*}{\textbf{No.}}&\multirow{2}{*}{\textbf{Synopsis}}&\multirow{2}{*}{\textbf{Category}}&\multicolumn{2}{|c|}{\textbf{Processor}}&\multicolumn{3}{|c|}{\textbf{Universal property-based method}}\\
            \cline{4-8}
            \textbf{}&\textbf{}&\textbf{}&\textbf{RIDECORE}&\textbf{BIRISCV}&\textbf{$\boldsymbol{SQED}$}&\textbf{$\boldsymbol{S^2QED}$}&\textbf{$\boldsymbol{TIUP}$}\\
            \hline
            $a01$&No implementation of divide-by-zero exception (Return $0xffffffff$)&single&\scalebox{0.75}{\usym{2613}}&\checkmark&\scalebox{0.75}{\usym{2613}}&\scalebox{0.75}{\usym{2613}}&\checkmark\\
            $a02$&No implementation of the division-modulo execution unit&single&\checkmark&\scalebox{0.75}{\usym{2613}}&\scalebox{0.75}{\usym{2613}}&\scalebox{0.75}{\usym{2613}}&\checkmark\\
            $a03$&Register target redirection&multiple&\checkmark&\checkmark&\checkmark&\checkmark&\checkmark\\
            $a04$&Register source redirection&multiple&\checkmark&\checkmark&\checkmark&\checkmark&\checkmark\\
            $a05$&Incorrect unsigned operand less-than compare&single&\checkmark&\checkmark&\scalebox{0.75}{\usym{2613}}&\scalebox{0.75}{\usym{2613}}&\checkmark\\
            $a06$&GPR0 can be assigned&multiple&\checkmark&\checkmark&\checkmark&\checkmark&\checkmark\\
            $a07$&Incorrect instruction fetched after dispatch stall&multiple&\checkmark&\checkmark&\checkmark&\checkmark&\checkmark\\
            $a08$&Incorrect instruction fetched after an LSU stall&multiple&\checkmark&\checkmark&\checkmark&\checkmark&\checkmark\\
            $a09$&One of the buggy RS-m entries corrupted the MULH/MULHU instruction&multiple&\checkmark&\scalebox{0.75}{\usym{2613}}&\checkmark&\checkmark&\checkmark\\
            $a10$&Erroneous branch addresses&single&\checkmark&\checkmark&\scalebox{0.75}{\usym{2613}}&\scalebox{0.75}{\usym{2613}}&\checkmark\\
            $a11$&Erroneous branch directions&single&\checkmark&\checkmark&\scalebox{0.75}{\usym{2613}}&\scalebox{0.75}{\usym{2613}}&\checkmark\\
            $a12$&Error in decoding next instruction's operand&single&\checkmark&\checkmark&\scalebox{0.75}{\usym{2613}}&\scalebox{0.75}{\usym{2613}}&\checkmark\\
            $a13$&Processor incorrectly decodes the next instruction to a NOP instruction&multiple&\checkmark&\checkmark&\checkmark&\checkmark&\checkmark\\
            $a14$&The value of the next register read is corrupted to all 0's&multiple&\checkmark&\checkmark&\checkmark&\checkmark&\checkmark\\
            $a15$&Erroneous speculative instruction aren't flushed&single&\checkmark&\checkmark&\scalebox{0.75}{\usym{2613}}&\scalebox{0.75}{\usym{2613}}&\checkmark\\
            $a16$&Unsigned multiply operand converts to signed&single&\checkmark&\checkmark&\scalebox{0.75}{\usym{2613}}&\scalebox{0.75}{\usym{2613}}&\checkmark\\
            $a17$&Source operand is misidentified as 0&multiple&\checkmark&\checkmark&\checkmark&\checkmark&\scalebox{0.75}{\usym{2613}}\\
            $a18$&ALU opcode does not match with the actual circuit&single&\checkmark&\checkmark&\scalebox{0.75}{\usym{2613}}&\scalebox{0.75}{\usym{2613}}&\checkmark\\
            $a19$&Error in determining whether instructions in decode queue have been popped&multiple&\scalebox{0.75}{\usym{2613}}&\checkmark&\checkmark&\checkmark&\checkmark\\
            $a20$&Logical error in fetch instruction valid signal&multiple&\checkmark&\checkmark&\checkmark&\checkmark&\checkmark\\
            \hline
            \multicolumn{5}{|c|}{\textbf{Detect single-instruction anomalies}}&\scalebox{0.75}{\usym{2613}}&\scalebox{0.75}{\usym{2613}}&\checkmark\\
            \hline
            \multicolumn{5}{|c|}{\textbf{Detect multiple-instruction anomalies}}&\checkmark&\checkmark&\checkmark\\
            \hline
            \multicolumn{5}{|c|}{\textbf{Runtime (with anomalies) [min, max]}}&[$<60s,>1.5h$]&[$<60s,>1.5h$]&[$<90s,<988s$]\\
            \hline
            \multicolumn{5}{|c|}{\textbf{Runtime (without anomalies)}}&Timeout&Timeout&$<988s$\\
            \hline
            \multicolumn{5}{|c|}{\textbf{Counterexample length ([min, max] instructions)}}&[$2, 14$]&[$2, 7$]&[$3, >14$]\\
            \hline
        \end{tabular}
        }
        \label{tab: anomaly}
    \end{center}
\end{table*}

\section{Conclusion}\label{sec: conclusion}
In this paper, we introduce $TIUP$, an end-to-end approach that employs tautologies as universal properties for formal verification of processors. $TIUP$ enables engineers to verify processors at the architectural level without the need for in-depth analysis of the microarchitecture. By utilizing a set of universal properties, $TIUP$ prevents the issues of false positives when using a single universal property and lets the verification problem be divided into a set of sub-problems, each of which is characterized through an abstract specification generated by a tautology. The experiment demonstrates the effective detection of anomalies in both in-order and out-of-order processors by $TIUP$.

\section*{\sc Acknowledgements}
This work was supported by Strategic Priority Research Program of Chinese Academy of Sciences under Grant No. XDC05020201. We also thank the anonymous reviewers for their feedback.


\bibliographystyle{IEEEtran}
\vspace{10pt}

\end{document}